\documentclass[reprint,prl,aps]{revtex4-1}

\usepackage{amsmath,amsfonts,amssymb,bm,dsfont,graphicx,mathtools}

\def\bp{\mathbf{p}}
\def\bu{\mathbf{u}}
\def\bx{\mathbf{x}}
\def\bF{\mathbf{F}}
\def\bX{\mathbf{X}}
\def\bPhi{\bm{\Phi}}
\def\grad{\mathbf{\nabla}}

\begin{document}

\title{Mechanics and morphology of proliferating cell collectives with self-inhibiting growth}

\author{Scott Weady$^{1}$, Bryce Palmer$^{1,2}$, Adam Lamson$^{1}$, Taeyoon Kim$^{3}$, Reza Farhadifar$^{1}$, Michael J. Shelley$^{1,4}$}

\email{sweady@flatironinstitute.org}

\affiliation{$^{1}$Center for Computational Biology, Flatiron Institute, New York NY 10010, USA}
\affiliation{$^{2}$Dept. of Mechanical Engineering, Michigan State University, East Lansing, MI 48824, USA}
\affiliation{$^{3}$Weldon School of Biomedical Engineering, Purdue University, West Lafayette, IN 47907, USA}
\affiliation{$^{4}$Courant Institute of Mathematical Sciences, New York University, New York, NY 10012, USA}

\date{\today}
\begin{abstract}
We study the dynamics of proliferating cell collectives whose microscopic constituents' growth is inhibited by macroscopic growth-induced stress. Discrete particle simulations of a growing collective show the emergence of concentric-ring patterns in cell size whose spatio-temporal structure is closely tied to the individual cell's stress response. Motivated by these observations, we derive a multiscale continuum theory whose parameters map directly to the discrete model. Analytical solutions of this theory show the concentric patterns arise from anisotropically accumulated resistance to growth over many cell cycles. This work shows how purely mechanical processes can affect the internal patterning and morphology of cell collectives, and provides a concise theoretical framework for connecting the micro- to macroscopic dynamics of proliferating matter.
\end{abstract}

\maketitle

The morphology of a cell collective is affected by factors across scales, from the mechanics and sensing of individuals to system-wide environmental conditions. At the local level, cell elasticity and geometry influence packing configurations \cite{Farhadifar2007,Staple2010,You2018}, intermittent phases of growth and motility drive concentric density variations \cite{Shimada2004,Liu2011}, and chemical cues give rise to internal patterning and interfacial instabilities \cite{Adler1966,Budrene1991,Budrene1995,Lu2022,Martinez-Calvo2022}. Global factors such as fluid flows \cite{Atis2019,Pearce2019,Benzi2022} or confinement \cite{Helmlinger1997,Yan2021,Amchin2022} can also modify growth rates and interfacial dynamics. Essential to all of these features is cell proliferation, in which individuals produce biomass from the consumption of chemical nutrients \cite{Hallatschek2023}. 

In contrast to conventional active matter systems, such as particle suspensions whose configuration is specified by spatial and orientational degrees of freedom, systems undergoing proliferation possess additional degrees of freedom that might include, for example, age or size \cite{Rubinow1968,Lebowitz1974,Wittmann2023}. Another distinguishing feature of proliferating systems are growth-induced stresses, which are particularly significant in cell collectives due to the inevitable many-body collisions between cells. These stresses not only cause local cellular rearrangements \cite{You2021}, but may also impact growth rates \cite{Delarue2016,Alric2022}. 

The influence of stress on growth and morphology has been analyzed in both discrete and continuum settings \cite{Shraiman2005,Gniewek2019,Schnyder2020,Carpenter2024,Ye2024}. Less attention has been paid to the coupled dynamics of growth-induced stress and microstructure. In this work, we study how stress-dependent growth affects the spatiotemporal dynamics and structure of proliferating colonies of elongated particles, such as bacteria or yeast. Many-particle simulations reveal concentric wave-like patterns in cell size whose space and time scales are determined by the sensitivity of individual cell growth to local mechanical stress. Based on this discrete model, we derive a continuum theory that couples the microscopic size degrees of freedom to the macroscopic mechanics, and find exact solutions in radially symmetric geometries. These solutions identify the origin of cell size patterning as the result of cumulative resistance to growth across generations.

\begin{figure}[b!]
\includegraphics[width=\linewidth]{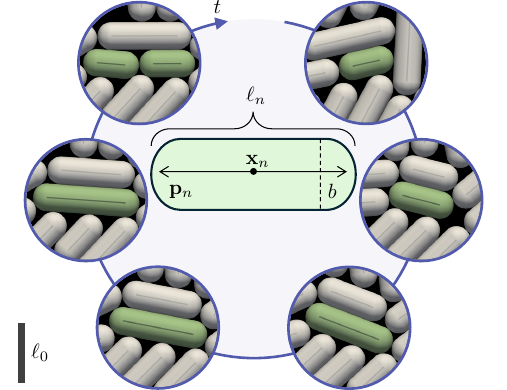}
    \centering
    \caption{Schematic of particle simulations. Each particle, indexed by $n = 1,\ldots,N(t)$, is a sphero-cylinder of constant diameter $b$. The particle configuration is described by a center of mass $\bx_n(t)$, orientation vector $\bp_n(t) = [\cos\theta_n(t),\sin\theta_n(t)]$, and length $\ell_n(t)$. As particles grow, they experience a compressive stress $\sigma^c_n$ along their principal axis due to contact with their neighbors which impedes their growth rate.}
    \label{fig:fig1}
\end{figure}

\begin{figure*}[t!]
    \centering
    \includegraphics[width=\linewidth]{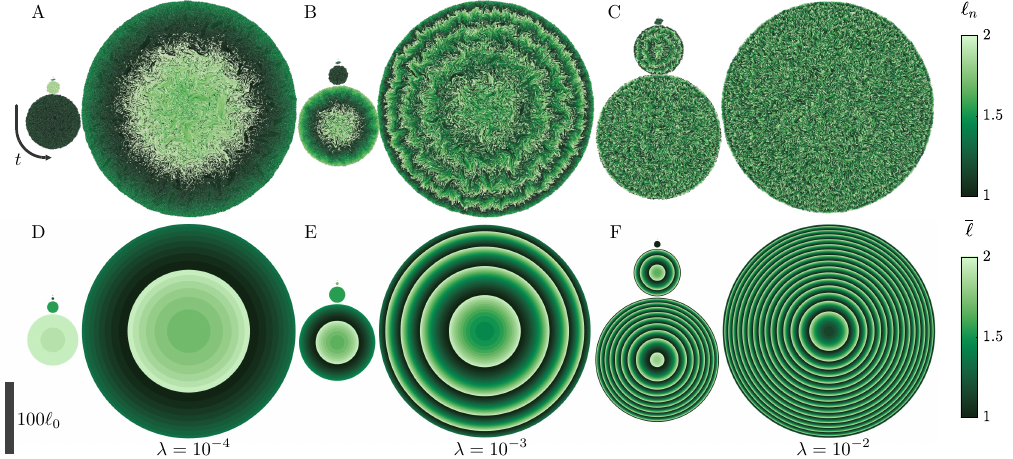}
    \caption{Pattern formation under stress-sensitive growth. Panels (A)-(C) show results of particle simulations with $\lambda = 10^{-4},10^{-3}$, and $10^{-2}$, respectively, each at five equally spaced points in time up to the time when the radius reaches $150\ell_0$, consisting of $\mathcal{O}(10^5)$ particles. The particles are colored by their length. As stress sensitivity increases, concentric rings form, and the characteristic wavelength of these rings decreases. Solutions of the continuum theory, (D)-(F), show similar pattern formation with a comparable wavelength. All parameters are in dimensionless form. Movies of these simulations are available as Supplementary Material.
    }
    \label{fig:fig2}
\end{figure*}

 Discrete models of cell proliferation are invaluable for their precise control of environmental parameters and mechanical responses. In such models, contact resolution is a common challenge owing to the typically dense packing of cell bodies. Here, we employ a contact resolution algorithm that enforces a no-overlap constraint by means of a nonlinear convex optimization problem. This method builds on contact resolution algorithms originally developed for granular matter and computer graphics \cite{steward_first_lcp_1996,anitescu_portra_time_stepping_2002,mazhar_apgd_2015,nvidia_nonsmooth_newton_2019} that have recently been adapted to inertialess active matter systems \cite{yan_collision_stress_2019,alens_2022}. This constraint-based formulation avoids numerical stiffness associated with repulsive potentials, allowing for larger time steps at the cost of an iterative solution.
 
 We consider a collection of $N(t)$ sphero-cylindrical particles of constant cross-sectional diameter $b$ that grow from an initial length $\ell_0$ to length $2\ell_0$, at which point they divide into two particles of length $\ell_0$. For simplicity, we neglect variations in cell size after division \cite{Amir2014,Taheri-Araghi2015}. Each particle is characterized by its center of mass $\bx_n(t)$, orientation angle $\theta_n(t)$, and length $\ell_n(t)$, shown schematically in Fig. \ref{fig:fig1}. Particles within the ensemble move and rotate under the contact forces created by growth. Considering particles atop a two-dimensional substrate with constant friction factor $\xi$, the translational and angular velocities induced by the total contact forces $\bF_n^c$ and torques $T_n^c$ are $d\bx_n/dt = (1/\xi\ell_n) \bF_n^c$ and
 $d\theta_n/dt = (12/\xi \ell_n^3) T_n^c$, respectively. Here $\bF_n^c$ and $T_n^c$ are expressed in terms of non-negative Lagrange multipliers determined from the constraints that particles cannot overlap, non-contacting particles do not generate force, and collisions minimize the rate of frictional dissipation (see Supplementary Material). Between divisions, each particle grows according to $d\ell_n/dt = (\ell_n/\tau)\exp(-\lambda\sigma_{n}^c)$, where $\tau$ is a growth timescale, $\sigma_{n}^c$ is the compressive stress acting along the long axis of the particle, and $\lambda$ is the stress sensitivity parameter. For uninhibited growth ($\lambda = 0$ or $\sigma_{n}^c = 0$), individuals grow exponentially fast, which has been reported experimentally for several types of rod-like bacteria, though this is still debated \cite{Godin2010,Campos2014,Kar2021}. For stress-sensitive growth ($\lambda>0$), the growth rate decreases with the compressive stress, an effect which is also observed in live cells \cite{Minc2009,Delarue2016,Alric2022}. Like the forces and torques, the compressive stress $\sigma_{n}^c$ is expressed in terms of the contact Lagrange multipliers, which themselves depend on growth, resulting in nonlinear coupling between contact and growth when $\lambda \neq 0$. Further details on the formulation and its numerical implementation can be found in the Supplementary Material.

\begin{figure*}[t!]
    \centering
    \includegraphics[width=\linewidth]{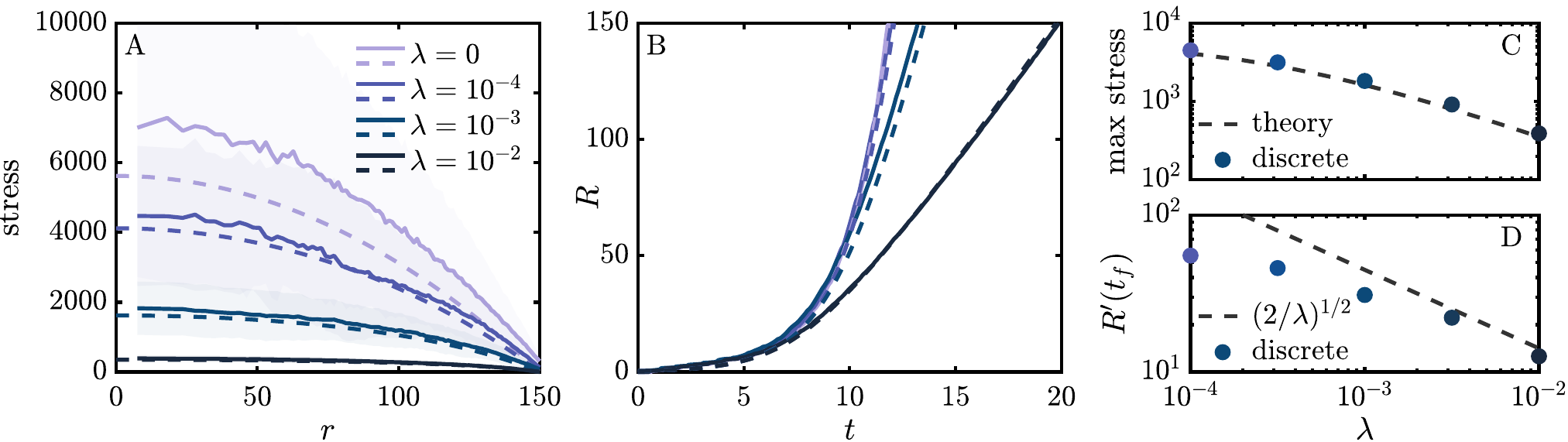}
    \caption{Mechanics of growth. (A) The radially averaged stress for both the discrete ($\langle\sigma_n^c\rangle$, solid lines) and continuum ($p$, dashed lines) models show that overall stress decreases with sensitivity. (B) The colony radius for both discrete and continuum simulations grows exponentially at early times, slowing to approximately linear growth at large times and sensitivities. (C) The maximum stress within the colony closely follows the scaling of the analytical solution. (D) A linear fit of the long-time rate of change of the colony radius indicates an approach to the asymptotic growth rate $\sim(2/\lambda)^{1/2}$.}
    \label{fig:fig3}
\end{figure*}

We non-dimensionalize the equations of motion by the cell length $\ell_0$, growth time scale $\tau$, and stress scale $\tau/\xi\ell_0^2$. The sole free parameter is the dimensionless growth sensitivity $\lambda' = (\tau/\xi\ell_0^2)\lambda$. The simulations begin with a single cell of unit length and diameter $b = \ell_0/2$, and run until the colony radius reaches $R = 150\ell_0$, corresponding to $\mathcal{O}(10^5)$ particles. Each panel in Fig. \ref{fig:fig2}(A-C) shows five snapshots, equally spaced in time, of simulations for three different values of $\lambda'$, with the particles colored by their length. As $\lambda'$ increases, concentric waves form about the colony center. Their characteristic wavelength decreases with $\lambda'$ up to a point where it approaches the particle scale and is difficult to distinguish by eye.

To gain insight into these patterns and their dynamic development, we formulate a continuum theory that describes both the collective macroscopic mechanics and the microscopic distribution of particle sizes. Let $\Psi(\bx,\ell,t)$ denote the number of particles with length $\ell\in[\ell_0,2\ell_0]$ in a control volume centered at $\bx\in\Omega(t)$, where $\Omega(t)$ is the colony domain. This distribution function satisfies a Smoluchowski equation in the material frame,
\begin{equation}
    \frac{D\Psi}{Dt} + \frac{\partial(\dot\ell\Psi)}{\partial \ell} = 0,\label{eq:dpsi/dt}
\end{equation}
where $D/Dt = \partial/\partial t + \bu\cdot\grad$ is the material derivative with the cell velocity field $\bu(\bx,t)$, and $\dot\ell(\bx,\ell,t) = (\ell/\tau)\exp(-\lambda p)$ is a growth flux, identical to the single particle growth rate in the discrete model with the compressive stress $\sigma^c$ replaced by an isotropic pressure $p(\bx,t)$. In this model, cell division is described by a boundary condition on $\Psi$ with respect to $\ell$. Specifically, the boundary condition balances the number density at the initial and division lengths,
$(\dot\ell\Psi)\big\rvert_{\ell=\ell_0} = 2(\dot\ell\Psi)\big\rvert_{\ell=2\ell_0}$,
where the factor of two comes from the assumption that cells divide into two daughter cells. A similar distribution with a division boundary condition was originally formulated to describe cell maturation \cite{Rubinow1968}, and was recently used to describe particle length distributions in the context of dynamic density functional theory \cite{Wittmann2023}.

The macroscopic mechanics are described by a Darcy-like model, which has been applied to collective cell growth in many contexts \cite{Greenspan1976,Roose2007,Lowengrub2009,Wu2013,Martinez-Calvo2022,Ye2024}. In the present context, this model arises as the minimizer of frictional dissipation between the particles and the substrate subject to a volumetric growth constraint, analogous to the principle satisfied by the particle simulations (see Supplementary Material). Specifically, the constitutive equations are
\begin{align}
    \xi\bu + \grad p = \bm{0},\quad
    \grad\cdot\bu = g,\quad&\text{for}~\bx\in\Omega(t),\label{eq:darcy}\\
    p(\bX,t) = 0,\quad \partial_t\bX = \bu(\bX,t),\quad&\text{for}~ \bX\in\partial\Omega(t).\label{eq:darcy_bc}
\end{align}
Here $\xi$ is a constant friction factor as in the discrete model and $g = [\dot\ell]_\ell/[\ell]_\ell = (1/\tau)e^{-\lambda p}$ is the volumetric growth rate with $[f]_\ell := \int_{\ell_0}^{2\ell_0} f\Psi ~ d\ell$. We non-dimensionalize the continuum equations using the same characteristic scales as before. The sole free parameter is again the dimensionless growth sensitivity $\lambda' = (\tau/\xi\ell_0^2)\lambda$, which exactly corresponds to that in the discrete model. In the remainder of this work, and in all figures, we assume all variables are dimensionless and omit primes. Further details on the model and its derivation can be found in the Supplementary Material.

Because $g$ does not depend on $\Psi$, the system of Eqs. (\ref{eq:darcy})-(\ref{eq:darcy_bc}) can be solved independent of the Smoluchowski equation (\ref{eq:dpsi/dt}), after which Eq. (\ref{eq:dpsi/dt}) can be solved along spatial and length characteristics. From this solution, given a Dirac delta initial condition ${\Psi_0(\bx,\ell) = \delta(\ell-1)}$, the average cell length $\overline{\ell} := [\ell]_\ell/[1]_\ell$ can be computed explicitly as
\begin{equation}
    \overline\ell(\bx,t) = 2^{{\rm frac}[\int_0^t g(\bPhi^{-1}(\bx;t'),t')/\ln 2 ~ dt']},\label{eq:lbar}
\end{equation}
where ${\rm frac}[z] \in [0,1)$ is the non-integer part of $z\geq 0$ and $\bPhi(\bx_0;t):\Omega(0)\mapsto\Omega(t)$ is the Lagrangian flow map. For uninhibited growth, this gives the homogeneous solution $\overline\ell(\bx,t) = 2^{{\rm frac}[t/\ln 2]}$, implying cells divide synchronously at time $t = n\ln 2$ for integer $n$, as expected. In contrast, for stress-sensitive growth inhomogeneities may arise through anisotropically accumulated resistance along characteristics. Notably, inhomogeneity is not necessarily a result of instability.

\begin{figure*}[t!]
\centering
\includegraphics[width=\linewidth]{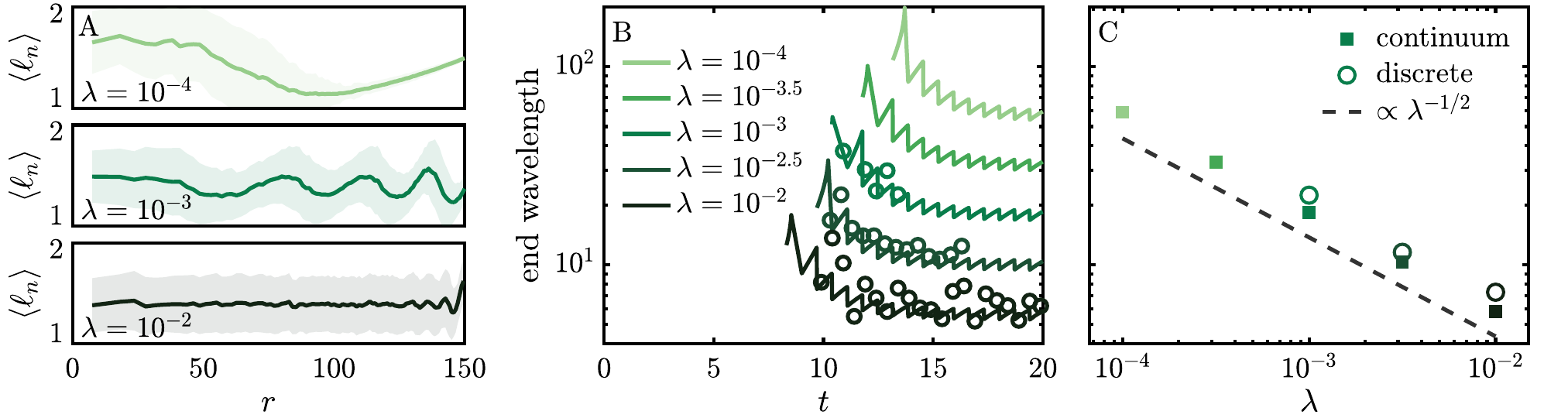}
\caption{Dynamics of size inhomogeneity. (A) The radially averaged particle length $\langle\ell_n\rangle$ in discrete simulations, binned in annuli of equal area, shows a wave-like signal, particularly at large values of $r$. The shaded region denotes one standard deviation from the mean. (B) The time evolution of the final wavelength in both continuum (solid lines) and discrete (open circles) simulations decays over time, approaching a constant value at late times. (C) The final wavelength demonstrates a $\lambda^{-1/2}$ scaling as suggested by a rescaling of the continuum pressure and characteristics.}\label{fig:fig4}
\end{figure*}

Motivated by the particle simulations, we consider radially symmetric colonies of radius $R(t)$ with $R(0) = \sqrt{b/\ell_0\pi}$, the initial size of which is chosen so that the initial area of the colony is approximately that of a single particle. In this geometry, Eqs. (\ref{eq:darcy})-(\ref{eq:darcy_bc}) can be solved analytically in $r = |\bx| \leq R(t)$ for the pressure $p$ and radial velocity $u$. For $\lambda = 0$ the solution is $p(r) = (R^2-r^2)/4$ and $u(r) = r/2$, while for $\lambda\neq 0$ the solution is
\begin{equation}
    p(r) = \frac{2}{\lambda}\ln\left(\frac{1}{8c}-c\lambda r^2\right),~ u(r) = \frac{4cr}{1/8c - c\lambda r^2},\label{eq:analytic}
\end{equation}
where $c(R) = [(1 + \lambda R^2/2)^{1/2}-1]/2\lambda R^2$. Figure \ref{fig:fig2}(D-F) shows the average cell length $\overline{\ell}$ from the continuum theory for the same parameters as the discrete simulations, without fitting. Similar concentric waves form whose wavelength closely follows that of the discrete simulations. At the largest sensitivity $\lambda = 10^{-2}$ [Fig. \ref{fig:fig2}(F)], the wavelength is roughly $5\ell_0$, suggesting a wave-like structure underlies the seemingly random distribution in the discrete simulation of (C).

The analytical solution given in Eq. (\ref{eq:analytic}) provides various predictions for the macroscopic dynamics and their dependence on $\lambda$. Figure \ref{fig:fig3}(A) shows the radially averaged stress from the discrete simulations $\langle\sigma_n^c\rangle$ (solid), binned in annuli of equal area, along with the analytical solution for $p$ (dashed). To reduce noise, the discrete stress is averaged over a time interval during which the colony grows by less than one particle length. As $\lambda$ increases, we find the overall stress decreases, with an order of magnitude difference between the $\lambda = 0$ and $\lambda = 10^{-2}$ cases. The continuum stress is typically lower than the discrete stress, likely due to the coarse-graining assumption that particles are isotropic in orientation. As a consequence, we do not expect exact agreement; however, we do anticipate equivalent scaling with $\lambda$. Indeed, the maximum pressure, shown in Fig. \ref{fig:fig3}(C), which is always located at the center of the colony, shows great agreement with the scaling of the analytical solution.

Figure \ref{fig:fig3}(B) shows the colony radius $R(t)$ over time for discrete (solid) and continuum (dashed) simulations for various $\lambda$. At early times the growth rates are comparable and show an exponential trend. The curves begin to diverge around $t \approx 5$, with larger values of $\lambda$ corresponding to slower growth. The continuum theory is in excellent agreement with the discrete model, showing that, while the model may differ at instantaneous points in time, as in Fig. \ref{fig:fig3}(A), it provides an accurate description of the system's dynamical evolution. From the continuum theory, for $\lambda = 0$ we have $R(t) = R_0e^{t/2}$. For any non-zero value of $\lambda$, however, taking $\lambda^{1/2}R\gg 0$ we instead find linear asymptotic scaling with time ${R(t) \sim (2/\lambda)^{1/2}t}$. Figure \ref{fig:fig3}(D) shows the rate of change $R'(t_f)$ at the final time, determined by a linear fit. At larger values of $\lambda$ we see an approach to the asymptotic slope, though it does not appear to be in the asymptotic regime for the lower values shown here.

To characterize the emergent waves, we consider the radially averaged length distribution $\langle \ell_n \rangle$ from the particle simulations, shown in Fig. \ref{fig:fig4}(A). For all non-zero sensitivities a wave-like signal is identified. The wavelength appears to decrease with the distance from the center of the colony $r$, and generally has a stronger signal at large distances. A quantitative characterization is given by the wavelength between the last two peaks, whose evolution in both discrete and continuum simulations is shown in Fig. \ref{fig:fig4}(B). The time of appearance of the first wave tends to decrease with $\lambda$, as indicated by the lack of data at early times. Once a well-defined wave has formed, its wavelength decreases over time, approaching an asymptotic value which scales with $\lambda^{-1/2}$ as shown in Fig. \ref{fig:fig4}(C). This scaling can be understood through the solution for the average length (\ref{eq:lbar}). Rescaling space as $\tilde r = \lambda^{1/2} r$ makes the growth rate $g(\tilde r)$ independent of $\lambda$. Because the colony grows linearly with rate $\propto\lambda^{-1/2}$ at late times, this suggests the integral in Eq. (\ref{eq:lbar}) approaches the same value for characteristics $\hat r_\lambda(t)$ with initial position $\hat r_\lambda(0) \propto \lambda^{1/2}$.

This work shows through large-scale particle simulations and a corresponding continuum theory how stress-sensitive growth affects the spatiotemporal dynamics of cell size in proliferating cell collectives. Concentric ring morphologies like those reported here are common and often form through chemical means and/or periodic processes \cite{Shimada2004,Liu2011,Liu2021}. In contrast, the patterns we observe arise from a constant process of mechanical resistance. In natural settings, such effects would be compounded by other growth-limiting factors such as access to nutrients, and controlled measurements may be needed to assess their relative importance to specific organisms. 

The continuum theory is generic and its predictions might apply to other cell collectives such as tissues, though material elasticity is likely an important factor that we do not address here \cite{Goriely2017}. Nonetheless, similar patterns have been observed in simulations of a vertex model of growing epithelial layers \cite{Carpenter2024}. A prediction of our model is that colonies of the same size can reduce internal pressure through stress-sensitivity, and organisms might exploit this property to optimize the balance of growth and stress. The predicted linear growth of the colony for non-zero sensitivities is consistent with experimental observations \cite{Hallatschek2007} and can be used to estimate the model parameter $\lambda$. Using data from \cite{Hallatschek2007}, for {\em E. Coli} the growth timescale is $\tau \approx 54{\rm min}$, the initial cell length is approximately $\ell_0 \approx 1\mu{\rm m}$, and the colony radial velocity is $dR/dt \approx 1.3\times 10^{-2} {\mu \rm m}/{\rm s}$, yielding the dimensionless sensitivity $\lambda \approx 1.1\times10^{-3}$, which is in the range of values where the mechanical effects discussed here are relevant.

The Smoluchowski equation provides an explicit coupling between the micro- and macro-scales, and its solution in terms of characteristics provides an analytical description of inheritance, or how the cumulative stress of a cell line affects the size and growth of subsequent generations. Within this context, other inherited variables, such as genes and mutations thereof, could be incorporated into the model and coupled to growth. Finally, while not discussed here, the discrete simulations exhibit rich orientational structure, and future work will explore the dependence of this structure on growth sensitivity.

\vspace{0.125in}

\begin{acknowledgments}
S.W. and B.P. contributed equally to this work. We thank Lev Bershadsky, Gilles Francfort, and Min Wu for engaging discussions. M.J.S. acknowledges support from NSF grant DMR-2004469. The computations in this work were performed at facilities supported by the Scientific Computing Core at the Flatiron Institute, a division of the Simons Foundation.
\end{acknowledgments}

\bibliography{refs}
\bibliographystyle{apsrev4-1}

\end{document}